\documentclass{aa}  

\usepackage{graphicx}
\usepackage{txfonts}
\usepackage{booktabs}
\usepackage{gensymb}
\usepackage{caption}
\usepackage{xtab}
\usepackage{siunitx}
\usepackage{hyperref}
\usepackage{natbib}
\usepackage{placeins}
\usepackage{caption}
\usepackage{float}
\hypersetup{
    colorlinks,
    citecolor=blue,
    filecolor=magenta,      
    linkcolor=blue,
    }

\newcommand{\plus}{$^{+}$}

\titlerunning{HCN/HCO\plus as a diagnostic tool in nearby galaxies}
\begin{document}

   \title{A multi-scale investigation into the diagnostic potential of the HCN/HCO\plus ratio for AGN and starburst activity in nearby galaxies}

   \subtitle{}

   \author{
J. Butterworth\inst{\ref{inst.Leiden}} 
\and S. Viti \inst{\ref{inst.Leiden},\ref{inst.TRA},\ref{inst.UCL}}
\and Y. Wang \inst{\ref{inst.Leiden}}
          }
\institute{\label{inst.Leiden}Leiden Observatory, Leiden University, PO Box 9513, NL-2300 RA Leiden, the Netherlands \\ \email{butterworth@strw.leidenuniv.nl}
\and\label{inst.TRA}Transdisciplinary Research Area (TRA) `Matter'/Argelander-Institut für Astronomie, University of Bonn
\and\label{inst.UCL}Physics and Astronomy, University College London, UK
}

   \date{Received ...; accepted ...}

 
  \abstract
{The identification of active galactic nuclei (AGN) and starburst (SB) regions in galaxies is crucial for understanding the role of various physical processes in galaxy evolution. Molecular line ratios, such as the HCN/HCO$^+$ ratio, have been proposed as potential tracers of these distinct environments.}
{This paper aims to assess the reliability of the HCN/HCO$^+$ ratio, from J = 1-0 to J = 4-3 transitions, as a diagnostic tool for differentiating AGN and SB activity across a diverse sample of nearby galaxies. We focus on evaluating the effect of spatial resolution on the robustness of these ratios and investigate the underlying physical conditions that drive observed variations.}
{We compile observations of HCN and HCO$^+$ lines across multiple J transitions from various sources, covering different galaxy types, including Seyferts, starbursts, and (ultra-)luminous infrared galaxies (U/LIRGs). The observations span spatial scales from cloud-sized regions (tens of parsecs) to kiloparsec scales. We analyse the behaviour of these ratios at varying resolutions and employ non-LTE radiative transfer models to infer the physical conditions that drive the observed ratios.}
{We find that the HCN/HCO$^+$ ratio from higher J transitions (e.g. J = 4-3) can differentiate between AGN and SB activity when observed at high spatial resolution ($< 100$ pc). This distinction occurs around unity, with HCN/HCO $<$ 1 being observed in SB-dominated and $>$ 1 in AGN-dominated regions. However, at lower resolutions, contamination from multiple emission sources and beam averaging effects destroy these distinctions. Radiative transfer modelling suggests that elevated HCN/HCO$^+$ ratios in AGN-dominated regions are largely driven by an enhancement in HCN abundance relative to HCO$^+$, likely due to high-temperature chemistry or increased excitation.}
{Our study confirms that the HCN/HCO$^+$ ratio, particularly of higher J transitions, can be a reliable tracer of AGN versus SB activity if observations are conducted at sufficiently high spatial resolution. However, caution must be exercised in interpreting these ratios at larger spatial scales due to contamination effects. Further high-resolution observations are needed to refine the conditions under which these ratios serve as reliable diagnostics.}

   \keywords{Interstellar medium (ISM): molecules, galaxies: active – Seyfert - starburst - ISM, astrochemistry.
               }

   \maketitle
%

\section{Introduction}
\label{sec:Intro}

Understanding the processes that drive galaxy evolution is a key challenge in astrophysics, and differentiating between the influence of active galactic nuclei (AGN) and starburst (SB) activity plays a crucial role in this endeavor. AGN and SB regions represent two dominant sources of energy in galaxies, each significantly impacting the physical and chemical environment of the interstellar medium (ISM). Identifying and characterizing these regions helps to unravel the mechanisms governing star formation, chemical enrichment, and feedback processes in galaxies.

Molecular line ratios have been used as a means of interpreting dense non-stellar gas environments, both galactic and extragalactic, for many years \citep{1998_Paglione,Kohno_2005,Izumi2016,2023_Imanishi}. For example, by comparing same-J molecular lines of molecules with similar intrinsic properties (i.e. critical density, $n_{crit}$, excitation temperature, $T_{ex}$) one can probe additional properties of that environment, such as cosmic ray ionisation rate and metallicity. These properties influence the excitation of these molecules within the ISM, and thus the intensity of their transitions. HCN and HCO\plus have been used in such a way for this very reason. As they possess transitions with high critical densities, these molecules trace the dense component of the gas within the ISM, such as in  Giant Molecular Clouds (GMCs). The excitation of these molecules within these regions are driven by collisions with H$_2$ and radiative excitation \citep{Shirley_2015}. \cite{Kohno2001} first proposed the `HCN diagram' which uses the HCN(1-0)/HCO\plus(1-0) alongside HCN(1-0)/ CO (1-0) ratios in order to differentiate between those galaxies with prominent X-ray dominated regions (XDRs) versus starburst galaxies \citep{Kohno2003,Kohno_2005}.
The HCN/CO ratio has been used in recent years as a tracer of the 'dense gas fraction` by making use of HCN's notably higher critical densities, assuming the use of same J transitions \citep[e.g.][]{Leroy2017}. The critical densities of the J=1-0 transitions of CO and HCN are $~10^3$ cm$^{-3}$ and $~10^{5}$ cm$^{-3}$,respectively \citep[][]{2001Flower,Shirley_2015}. The HCN/HCO\plus ratio, across multiple J-transitions have been proposed as a means of tracing AGN versus starburst activity \citep{Leonen2007,Izumi2013,Izumi2016,Butterworth2022} leading to varying results. HCN(1-0)/HCO\plus(1-0)  does show some variance between certain AGN and SB environments, typically with HCN(1-0)/HCO\plus(1-0)>1 in AGN- and <1 in SB-dominated regions \citep{Kohno_2005}. Some studies have found that this is not always the case; for example a statistical study conducted by \cite{Privon_2020} in which they surveyed a sample of 58 resolved, local luminous infrared galaxies (LIRGs) and ultra luminous infrared galaxies (ULIRGs), concluded that an enhancement in the HCN/HCO$^+$(1-0) line ratio could not be shown to be correlated to the AGN activity, and so HCN/HCO$^+$ ratios are not a dependable method for tracing AGN regions located within galaxies. In \cite{Butterworth2022} it was also found that observations of HCN(1-0)/HCO\plus(1-0) were not able to accurately differentiate between regions located in the AGN-dominated gas of circumnuclear disk (CND) of the nearby composite galaxy NGC 1068 and regions located in the galaxy's starburst ring. The higher J HCN(4-3)/HCO\plus(4-3) ratio, however, was able to differentiate these regions at both higher and lower resolutions. \cite{Butterworth2022} also investigated other possible AGN activity tracers, such as HCN(4-3)/CS(7-6) and HCN(4-3)/CS(2-1) and found that HCN(4-3)/CS(2-1) could clearly show a distinction between CND and SB regions at high resolution, but at low resolution the distinction was no longer clear. These results beg the question as to under which conditions and for which sources may these ratios be able to be used as a tracer. Furthermore, understanding the situations under which these ratios are effective may provide a better understanding of the underlying physical and chemical processes responsible for the observed variations in these ratios.

In terms of physical processes that may lead to the observed increased intensity of HCN lines with relation to lines of similar properties (such as same J HCO\plus lines), high temperature reactions have been proposed to be  responsible for an enhancement in HCN abundance observed in regions near AGN within nearby extragalactic sources compared to sources within high star-forming regions, such as SBs \citep{Izumi2016}. This scenario is one whereby neutral-neutral reactions with high reaction barriers are enhanced \citep{Harada_2010}, thus leading to the possible enhancement of HCN and the depletion of HCO$^+$ via newly available formation and destruction paths, respectively. Additionally, the higher temperatures could increase HCN excitation, relative to HCO$^+$ and CS, without necessarily changing their relative abundances \citep{Imanishi_2018}. The results from \cite{Butterworth2022} obtaining column density estimates of HCN, HCO\plus and CS within NGC 1068 would imply that an increased HCN abundance is the driving factor. It must be noted though that the HCN/HCO\plus ratio is dependent on multiple environmental properties, such as temperature and density, thus efforts to overcome degeneracies should be used when conducting analysis using these ratios \citep{Viti_2017,Butterworth2022}.


\par While in \cite{Butterworth2022} the HCN/CS ratio, in particular HCN(4-3)/CS(2-1), showed some promise as a tracer of AGN vs SB activity at high resolution there is currently a lack of common resolution observations of these lines across nearby galaxies for a homogeneous study to be conducted. As a result of the close proximity in frequency of same-J HCN and HCO\plus lines they are often covered in the same sideband of observations, resulting in a far larger available dataset in nearby galaxies at varying resolutions. We have therefore performed a multi-scale, multi-source investigation into the use of HCN/HCO\plus molecular line ratios in order to constrain the situations under which it is appropriate to use this ratio as a tracer of AGN or SB activity. This kind of study is necessary to overcome bias inherent to single source and/or single spatial resolution studies.
 In Sect. \ref{sec:Data} we provide a summary of the data collated for this study. Section \ref{sec:ratios} presents an analysis of various molecular ratios at both high and low spatial resolution. In Sect. \ref{sec:Propanalysis} we perform non-LTE analysis of the line intensities used for the ratios in Sect. \ref{sec:ratios}. We do this as a means of investigating whether this analysis, which incorporates multiple ratios at once can provide an understanding of the physical conditions leading to the trends observed in the intensity ratios. Moreover, whether the use of these models may more clearly distinguish between galactic types. We summarise our findings in Sect. \ref{sec:Conclusion}.

    

\section{Data}
\label{sec:Data}

In total this study examines a sample of 110 sources across 80 different galaxies, including multiple regions within the same galaxy. Our sample of galaxies are of numerous types (e.g. Seyfert, Starburst, (U)LIRG) and have been observed at various spatial resolutions. These variances of resolution, from sub-GMC scale to galactic scale, require that classification of each region. The type classification of sources in this paper depends on resolution: high-resolution observations reflect specific regions within a galaxy, while lower-resolution observations capture more global characteristics. The spatial resolutions of these observations range from cloud scale at $~25$ pc to galactic scale at $~20$ kpc, the details of these resolutions and how sources are grouped are described in Section \ref{sec:ratios}.  Classifications are taken either from the source of the data used, or in ambiguous cases from the NASA/IPAC Extragalactic Database (NED) \citep{NED}. Taking NGC 1068 as an example, as a composite galaxy, NGC1068 contains both an AGN and a starburst ring. With the use of high resolution interferometers, like the Atacama Large Millimeter/submillimeter Array (ALMA), the emission from these regions (and even from prominent sub-regions within) may be obtained independently from one another. The full list of sources, their respective observed transitions of HCN and HCO\plus along with their respective references are available from the CDS (see Data Availability). 

In the case of observations of the same object with the same instrument at the same resolution, which resulted in similar intensities only one result was included.
In order to properly examine the effect upon the observed intensity ratio by the observational spatial resolution, we accumulated data for each source, grouped by spatial resolution. For some sources, similar (but not the same) spatial resolution observations have been grouped in one category, for varying or the same transition.
Within each category, we then deconvolved the intensities by assuming the observation with the lowest resolution was the source size and using the equation
\begin{equation}
I_{Source}=\left(\frac{(\theta_{s}^2+\theta_{b}^2)}{\theta_{b}^2}\right)I_{Beam},
\end{equation}
where $I_{Source}$ is the source averaged peak intensity, $\theta_{s}$ is the source size (as defined by the lowest-resolution observation in the category), $\theta_{b}$ is the beam size, and $I_{Beam}$ is the original intensity as defined in the archive or paper. This is the same approach as taken in \cite{Butterworth2022} and based upon the approach used to handle beam dilution taken in \cite{Kamenetzky2011,Aladro2013,Aladro2015}.
The spatial resolution bins were chosen in an effort to group observations based upon likely physical scales, while also maintaining a reasonably similar spatial resolution variation (<2). As an example, a visual representation of the various scales covered by the observations utilised within this study are shown for the composite galaxy NGC~1068 in Appendix \ref{app:1068}.

\section{Molecular line ratios}
\label{sec:ratios}

In this section we shall investigate how each J-transition ratio behaves within each respective resolution group before then evaluating the effect of the spatial resolution of the observation upon these ratios. HCN and HCO$^+$ are tracers of dense gas, with their J=1-0 rotational transition lines possessing critical densities of $\sim$\SI{e5}{\per\centi\metre\cubed} and $\sim$\SI{5e4}{\per\centi\metre\cubed} \citep[at $\sim 10$K,][]{Shirley_2015}, respectively. Thus, any observed variances in same-J transitions of these similar molecules may be used to trace unique processes that may not be investigated in single molecule studies. \cite{Butterworth2022} found that HCN(4-3)/HCO$^+$(4-3) ratios had the ability to clearly distinguish between AGN and SB-dominated regions within NGC~1068. Therefore, in this section, we discuss each ratio in turn, with a particular focus on how well the ratio separates AGN-dominated regions from SB-dominated regions. In the cases where (U)LIRG sources are present they are also discussed. We take the typical size of GMCs (10s of parsecs), as a somewhat arbitrary  conceptual cut-off between what we refer to as `high' resolution and `low' resolution within this study.

\subsection{Cloud-scale and higher resolution}
\label{sec:Cloud_ratios}

\begin{figure*}
    \centering
    \includegraphics[width=\textwidth]{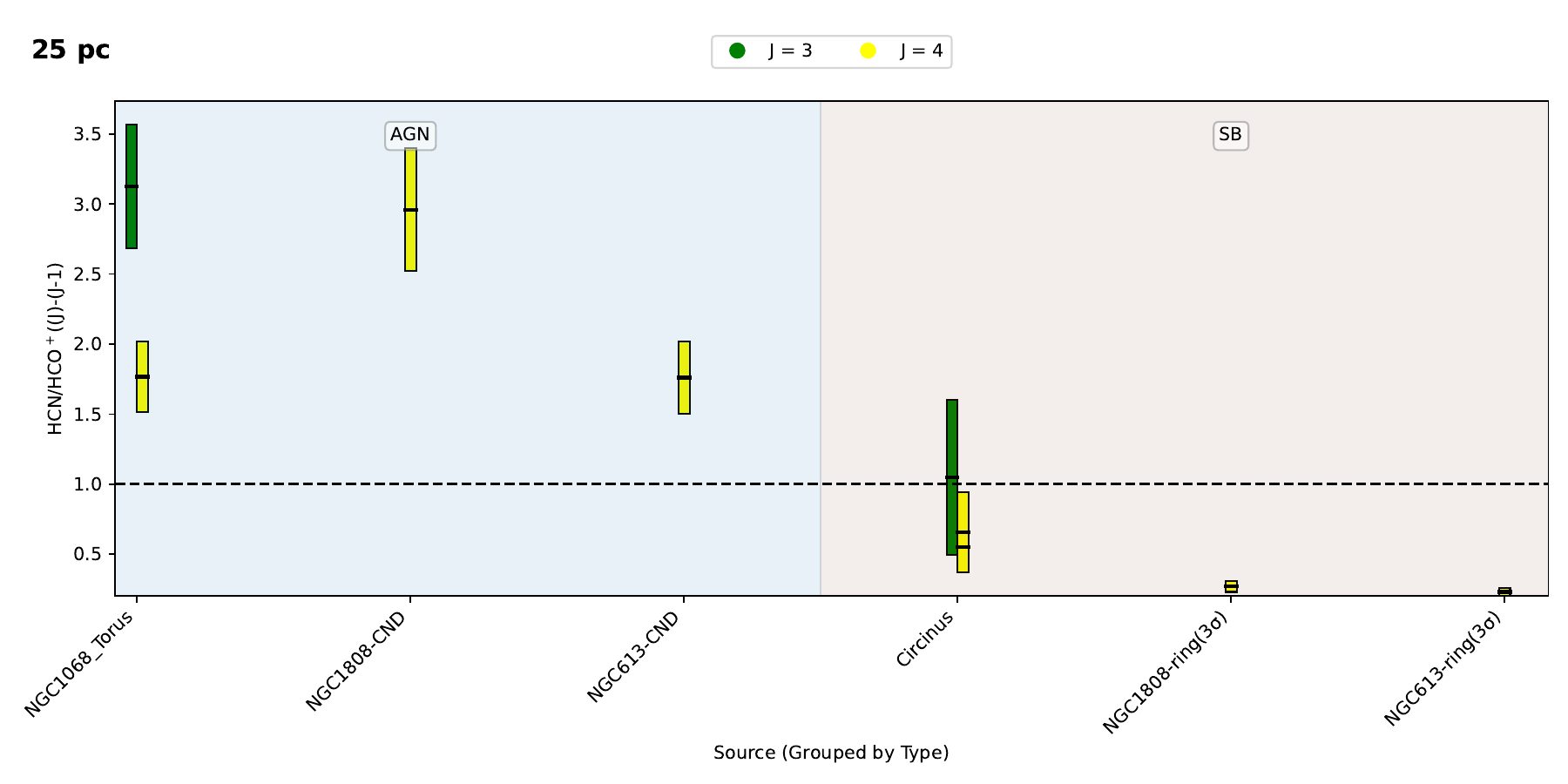}
    \caption{A bar swarm plot of the HCN/HCO\plus intensity ratios observed at $\sim25$ pc in regions of nearby galaxies. The colour of each bar displays the corresponding J-transition of the HCN(J-(J-1))/HCO\plus(J-(J-1)) presented.}
    \label{fig:highest_intens}
\end{figure*}

\begin{figure*}
    \centering
    \includegraphics[width=\textwidth]{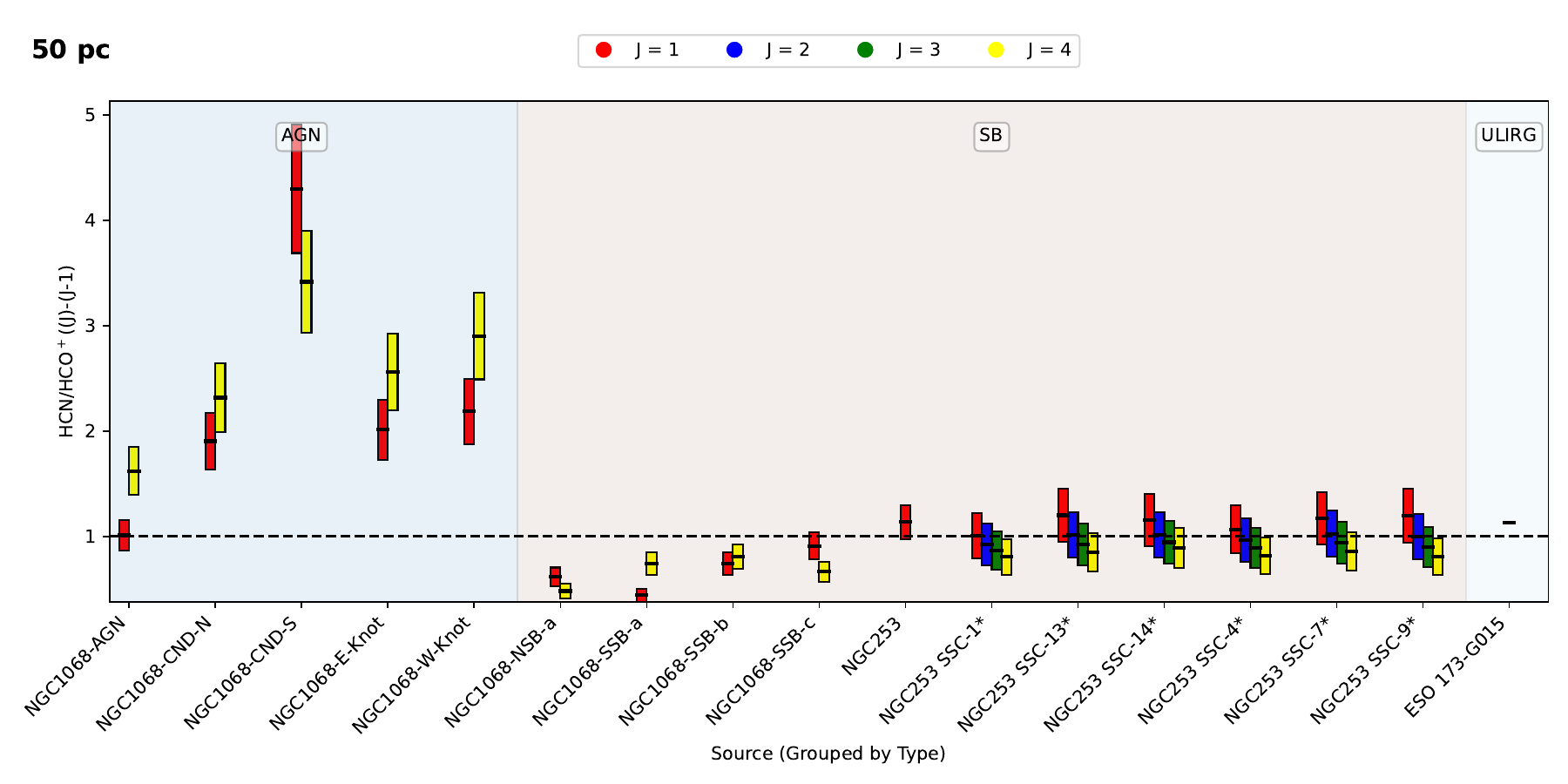}
    \caption{Same as Fig. \ref{fig:highest_intens} but $\sim50$ pc  resolution.}
    \label{fig:high_intens}
\end{figure*}
 Firstly, we shall now cover the molecular line ratios observed at $\lesssim50$pc. The intensities for these sources have been taken from studies that use a beam size aperture comparable to the size of each region (see Data availability). As such, sources such as NGC1808-ring are located within SB ring of NGC~1808 itself and is not an annuli encompassing the entire SB ring. The molecular line ratios from these sources are plotted in Figures \ref{fig:highest_intens}-\ref{fig:high_intens}. We focus first on the highest resolution group, up to 25 parsecs (Fig. \ref{fig:highest_intens}). Here we possess an equal number of AGN and Starburst sources though only in the high-J, J = (3-2) and J = (4-3) ratios.  As can be seen the HCN(4-3)/HCO\plus(4-3) ratio clearly separates the sources around unity, with AGN-dominated sources >1 and SB-dominated sources <1. This behaviour is consistent with the results of \cite{Butterworth2022}. HCN(3-2)/HCO\plus(3-2) however, observed in the torus of NGC~1068 and within the Circinus galaxy, may still trace AGN and Starburst activity though with the distinction no longer occurring around unity. 

Figure \ref{fig:high_intens} shows the behaviour of these ratios at up to 50 parsec scale. The designated names for each sub-region within each galaxy are as defined in their original papers, see Data Availability for the table containing the reference of each source. 
The HCN(4-3)/HCO\plus(4-3) ratio at this resolution again very clearly separates SB-dominated sources and AGN-dominated sources, with the AGN- and SB-dominated sources each separated above and below 1, respectively. The HCN(1-0)/HCO\plus(1-0) ratio notably does not separate the two types of regions at this resolution. The HCN(1-0)/HCO\plus(1-0) is shown to be larger than 1 when observed in regions of the central molecular zone (CMZ) of NGC~253 \citep{Meier_2015,2024Butterworth}. This is comparable to observations of this ratio at the AGN-position of NGC~1068 at $\sim50$pc that shows a value of unity. For each of these cases, there are possible explanations, at the AGN position there is a notable `cavity' of emission from HCN and HCO\plus, likely resulting from the ionized gas outflow known to exist in NGC~1068 \citep{2014_Garcia_Burillo,Huang_2022,2022_vollmer}. 
It is worthwhile to compare this observation to that of the higher resolution observation of the torus of NGC~1068, whereby high-J transitions are also able to trace between AGN and SB regions.
In the case of the CMZ of NGC~253, super star clusters (SSCs) are located within each of these observations; if high temperature chemistry resulting in an increased HCN abundance is a contributing factor then the high temperature of these regions may result in an increase of this ratio \citep{2020_Rico-Villas,2022_Rico_villas,2024Butterworth}. Regardless, these results concur with the results of \cite{Privon_2020} that the HCN(1-0)/HCO(1-0)\plus does not trace AGN activity unambiguously.

\subsection{Extended Resolution}

\begin{figure*}
    \centering
    \includegraphics[width=\linewidth]{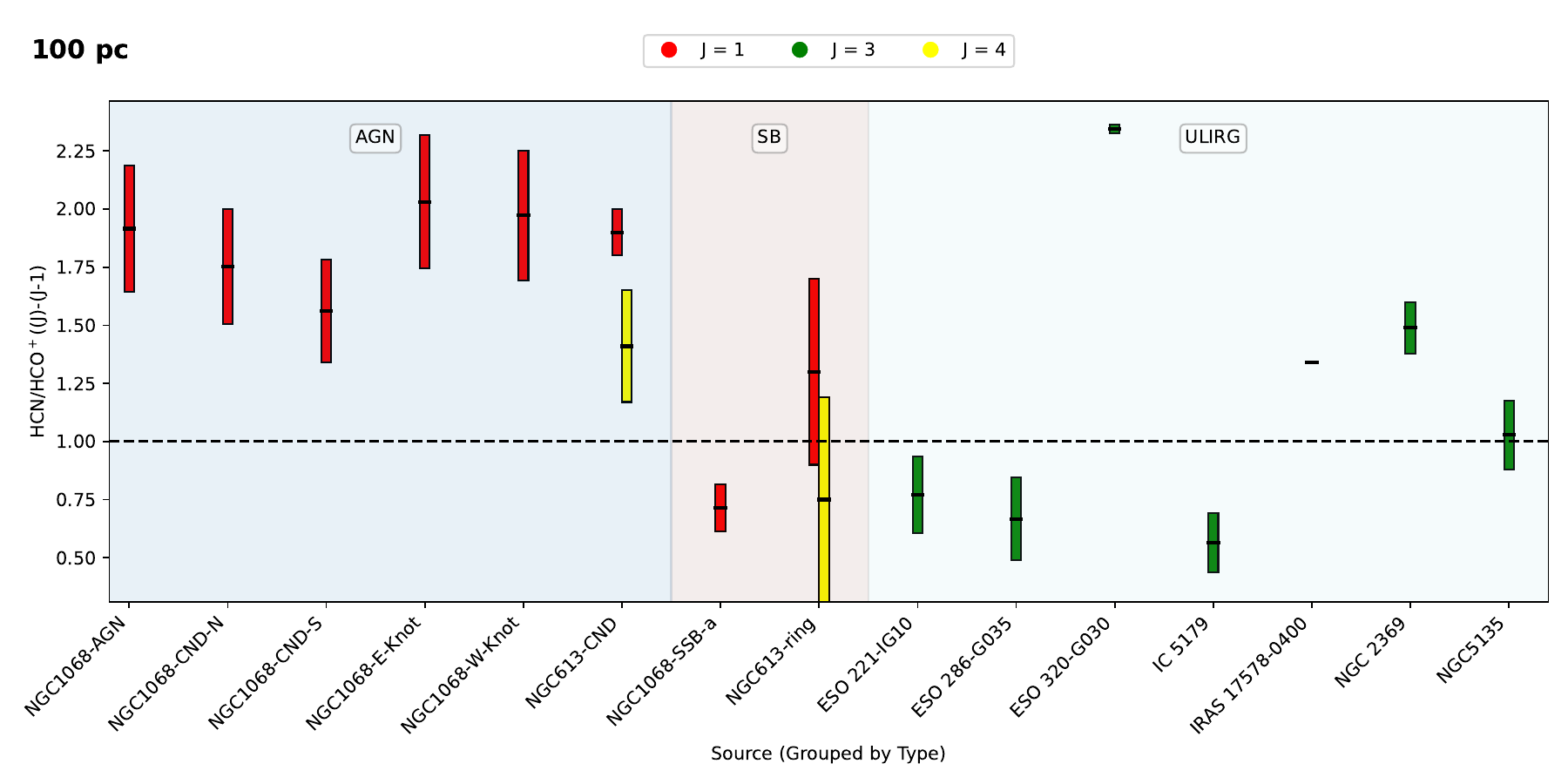}
    \caption{Same as Fig. \ref{fig:highest_intens} but $\sim100$ pc  resolution.}
    \label{fig:medium_intens}
\end{figure*}

\begin{figure*}
    \centering
    \includegraphics[width=\linewidth]{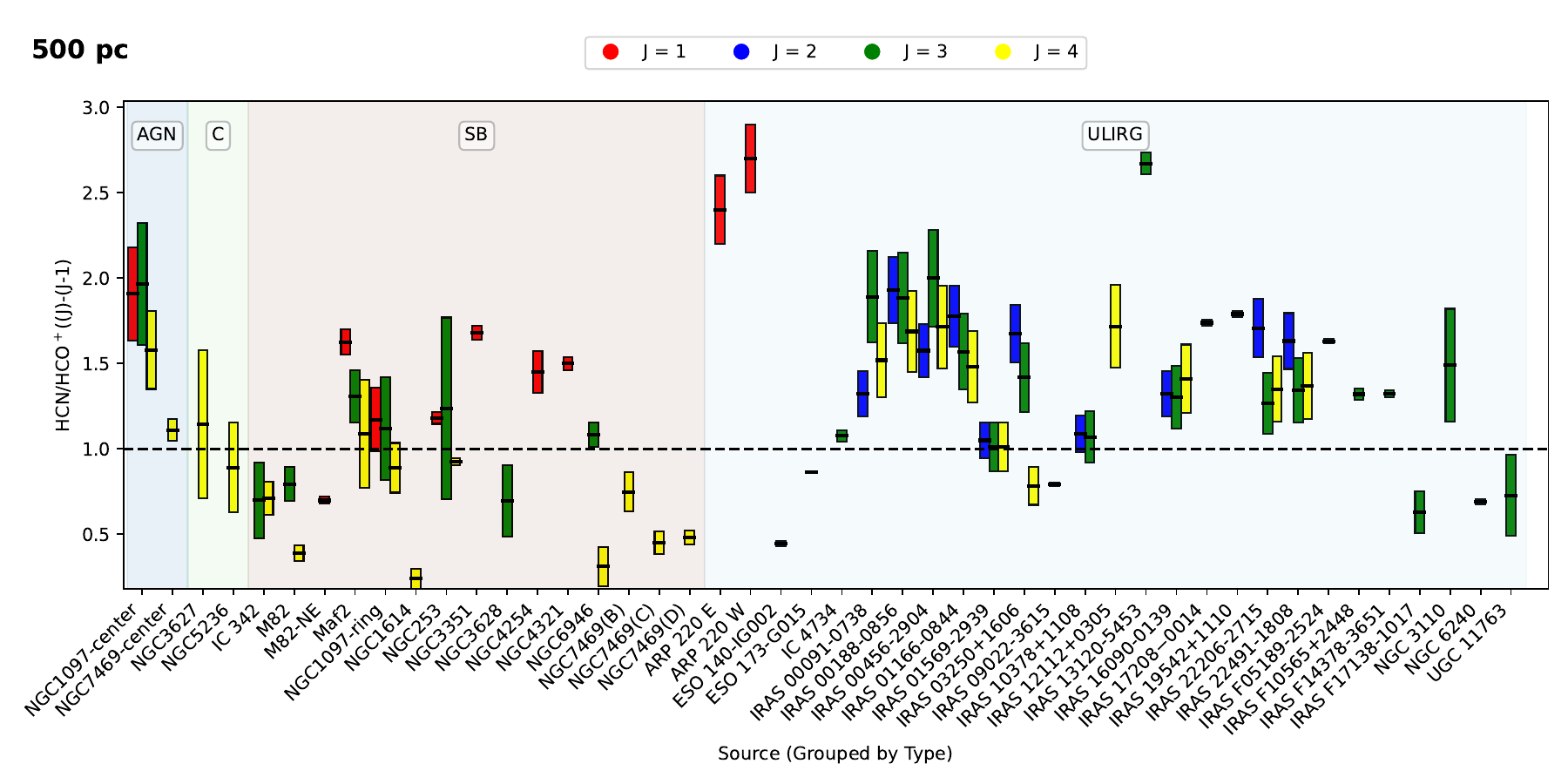}
    \caption{Same as Fig. \ref{fig:highest_intens} but $\sim500$ pc  resolution.}
    \label{fig:low_medium_intens}
\end{figure*}

At $\sim$100 pc and 500 pc scale, respectively, Figures \ref{fig:medium_intens} and \ref{fig:low_medium_intens} represent an intermediate group of resolutions between cloud-scale and galactic scale observations. In this regime, contamination from multiple emission sources must be considered, affecting the reliability of molecular line ratios as tracers. As such a new designation of composite (C) sources has been added to each plot. These are sources that contain both an AGN and a SB. This is particularly evident in Figure \ref{fig:low_medium_intens}, as at this resolution ($\sim 500$ pc) it is likely that the beam is no longer resolving a single source, such as a GMC. This is evident in the variance between both Figures \ref{fig:medium_intens} and \ref{fig:low_medium_intens}, and also how they compare to the higher resolution observations covered in Figures \ref{fig:highest_intens}-\ref{fig:high_intens}. At the $\sim 100$pc resolution shown in Figure \ref{fig:medium_intens} it can be seen that the regions of the CND of NGC~1068, that were also present in Figure \ref{fig:high_intens}, now show a greater homogeneity in the HCN(1-0)/HCO\plus(1-0) ratio at the AGN-position, no longer appearing to be located around 1. This is likely a result of the observation extending beyond the cavity observed around the AGN of NGC~1068 within the CND. 
Observations of the SB-ring of NGC~613 at this resolution appear to show that once again that HCN(1-0)/HCO\plus(1-0) does not clearly distinguish between AGN- and SB-dominated regions. 
Similarly, the HCN(4-3)/HCO\plus(4-3) ratio, though only observed in the CND and SB-ring of NGC613, at this resolution does not clearly trace either region. At the resolution covered by Figure \ref{fig:low_medium_intens} ($\sim\ $500 pc) significant contribution from multiple emission sources is likely. For example NGC~3351, though classified as a star-forming galaxy, shows a significantly high HCN(1-0)/HCO\plus(1-0) ratio of $\sim\ 1.7$  \cite{2017_Jimenex-donaire,2023_Garcia_Rodriguez}. This is consistent with \cite{Butterworth2022} in that ratios at high resolution that were once able to distinguish between AGN and SB activity, no longer maintain that diagnostic power at low resolution. This is also seen in overlap of ratio once lower resolution observations of sources are considered by other studies, such as \cite{Izumi2016}.

    

\subsection{Galactic-scale Resolution}

\begin{figure*}
    \centering
    \includegraphics[width=\linewidth]{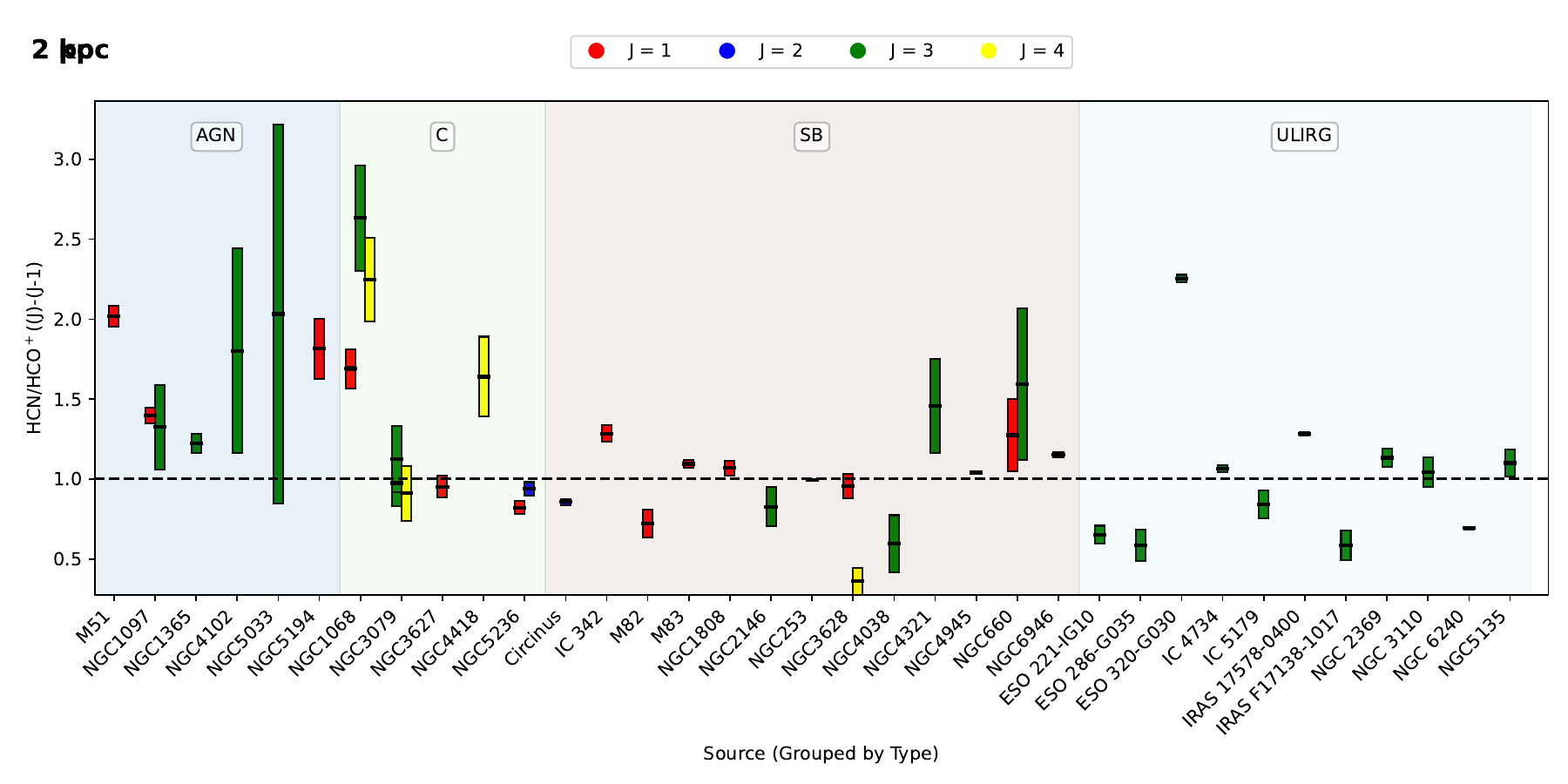}
    \caption{Same as Fig. \ref{fig:highest_intens} but $\sim200$ pc  resolution.}
    \label{fig:low_intens}
\end{figure*}

\begin{figure*}
    \centering
    \includegraphics[width=\linewidth]{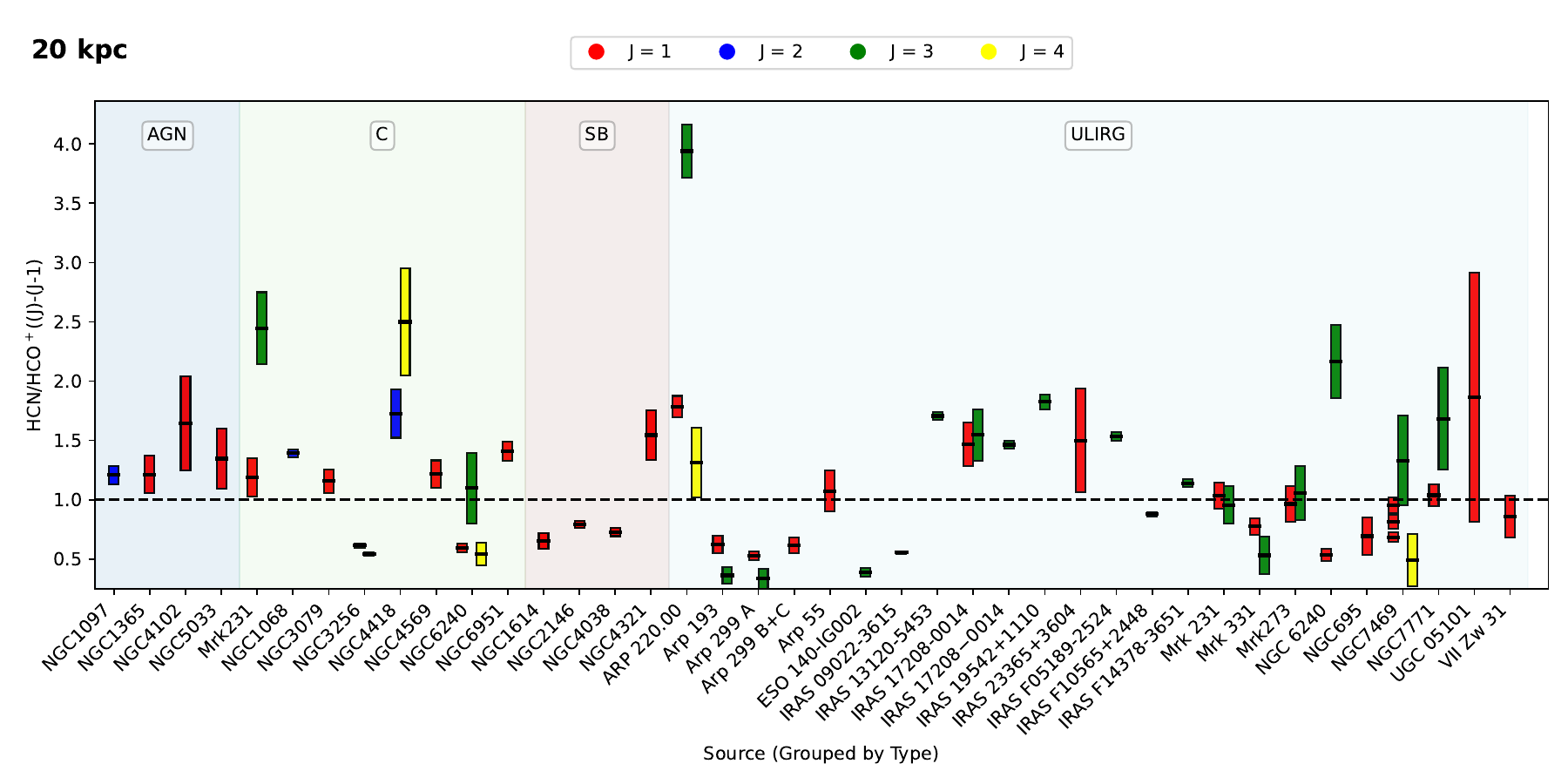}
    \caption{Same as Fig. \ref{fig:highest_intens} but $\sim20000$ pc  resolution.}
    \label{fig:lowest_intens}
\end{figure*}

Figures \ref{fig:low_intens} and \ref{fig:lowest_intens} contain primarily single dish low resolution observations of nearby galaxies. These observations, particularly in Fig. \ref{fig:lowest_intens}, begin to encompass entire galaxies, providing a general overview of global molecular gas distributions across the different galaxy types. As can be seen,  at both $\sim 2$kpc (Figure \ref{fig:low_intens}) and at $\sim 20$kpc \ref{fig:lowest_intens} resolution, the distinction between AGN-dominated and SB-dominated regions has mostly broken down, particularly the distinction around unity observed in the cloud-scale observations. This trend appears to occur regardless of the chosen transition. This behaviour is to be expected as the resolution of the observations increase the contamination of multiple emission regions, alongside with possible dilution effects beginning to take hold. Of some note perhaps is the fact that in general, non-ULIRG galaxies with a known AGN appear to show ratio values greater than 1 regardless of transition, with a minor exception within errors for NGC~5194. This distinction however is undermined perhaps with the greater variance observed in the remaining galaxy types. These results agree with the ambiguity previously observed in smaller sample studies attempting to distinguish between AGN and SB galaxies on these scales, such as in \cite{Kohno_2005}, \cite{Krips_2008} and \cite{2023_Israel}.



\section{Non-LTE Analysis}

\label{sec:Propanalysis}

As shown by  looking at the line intensity ratios, the high-J transitions of HCN and HCO\plus are able to distinguish between AGN and starburst activity, at high resolution. In this section we explore, via the use of radiative transfer modelling, whether this is the result of an enhancement of HCN abundance with respect to HCO\plus, as was the conclusion in \cite{Butterworth2022} and other studies \citep{Izumi2016,2023_Imanishi}.

 To determine the physical conditions from each observation, we followed a similar methodology of combining \texttt{SpectralRadex} and \texttt{UltraNest}, as defined in \citep{2021Holdship}, \citep{2023Huang} and \citep{2024Butterworth}. \texttt{SpectralRadex} employs the use of the radiative transfer modelling code \texttt{RADEX} \citep{Van_der_Tak_2007}. The physical conditions constrained within our modelling are: neutral gas density ($n_{\rm H2}$), kinetic temperature ($T_{\rm kin}$) and the total column density of both HCN and HCO\plus. The posterior probability distributions are obtained with the nested sampling Monte Carlo algorithm MLFriends \citep{ultranest16,ultranest19}. This algorithm has been integrated into our analysis by using the Python package
\texttt{UltraNest} \footnote{\url{https://johannesbuchner.github.io/UltraNest/}} \citep{ultranest21}. 
We make the assumption that the prior distributions of our parameters are uniform or log-uniform; these priors are given in Table \ref{tab:table_prior}. These priors were chosen to sample a physically reasonable parameter range: a gas density, $n_{\text{H}2}$, between $10^3$ and $10^7$cm$^{-3}$ covers the range of densities of gas in which we would expect HCN and HCO\plus, and this range also includes the critical densities of the transitions covered in this paper. The kinetic temperature range of $20-200$ K was dictated by the range of temperatures covered by the collisional data of both HCN and HCO\plus, as sourced from the Leiden Atomic and Molecular Database \citep[LAMDA][]{2020_hcop_collis,2020_LAMDA,2023_HCN_rates}. The column density ranges were chosen in an effort to cover the range of diffuse to dense regions possibly probed by such a study considering such varying spatial scales. We assume that the uncertainty on our measured intensities, and thus ratios, are Gaussian such that the likelihood is given by $P(\theta | d) \sim \exp(-\frac{1}{2}\chi^2),$ where $\chi^2$ is the chi-squared statistic between our measured intensities and the \texttt{RADEX} output for each set of physical parameters $\theta$. In order to obtain as many ratios as fitted parameters we only perform this modelling on sources that have at least observations of two discrete J-transitions of HCN and HCO\plus of the same source at the same resolution. We then fit the respective HCN(J-(J-1))/HCO\plus(J-(J-1)) ratios along with the respective excitation ratios (e.g. HCN(J-(J-1)/HCN(I-(I-1), where J$>$I) thus obtaining at least 4 ratios for each source-resolution pair.

This analysis offers an estimate of the average gas properties for each observation across our sources with multiple transitions, assuming that all the fitted transitions within resolution-source pair originate from the same gas component.

\begin{table}
  \centering
  \caption{Parameter space allowed as a prior for the nested sampling of the HCN and HCO\plus \texttt{RADEX} models across the various sources.}
  \label{tab:table_prior}
  \begin{tabular}{c|cc}
  \hline
    Variable  & Range & Distribution type\\
    \hline \hline
    Gas density, $n_{\rm H2}$ [cm\textsuperscript{-3}] & $10^{3}-10^{7}$ & Log-uniform\\
    Kinetic temperature, $T_{\rm kin}$ [K] & $20-200$ & Uniform\\
    $N$(HCN) [cm\textsuperscript{-2}] & $10^{12}-10^{17}$ & Log-uniform\\
    $N$(HCO\plus) [cm\textsuperscript{-2}] & $10^{12}-10^{17}$ & Log-uniform\\
    \hline
  \end{tabular}
\end{table}

\begin{figure*}
    \centering
    \includegraphics[width=\linewidth]{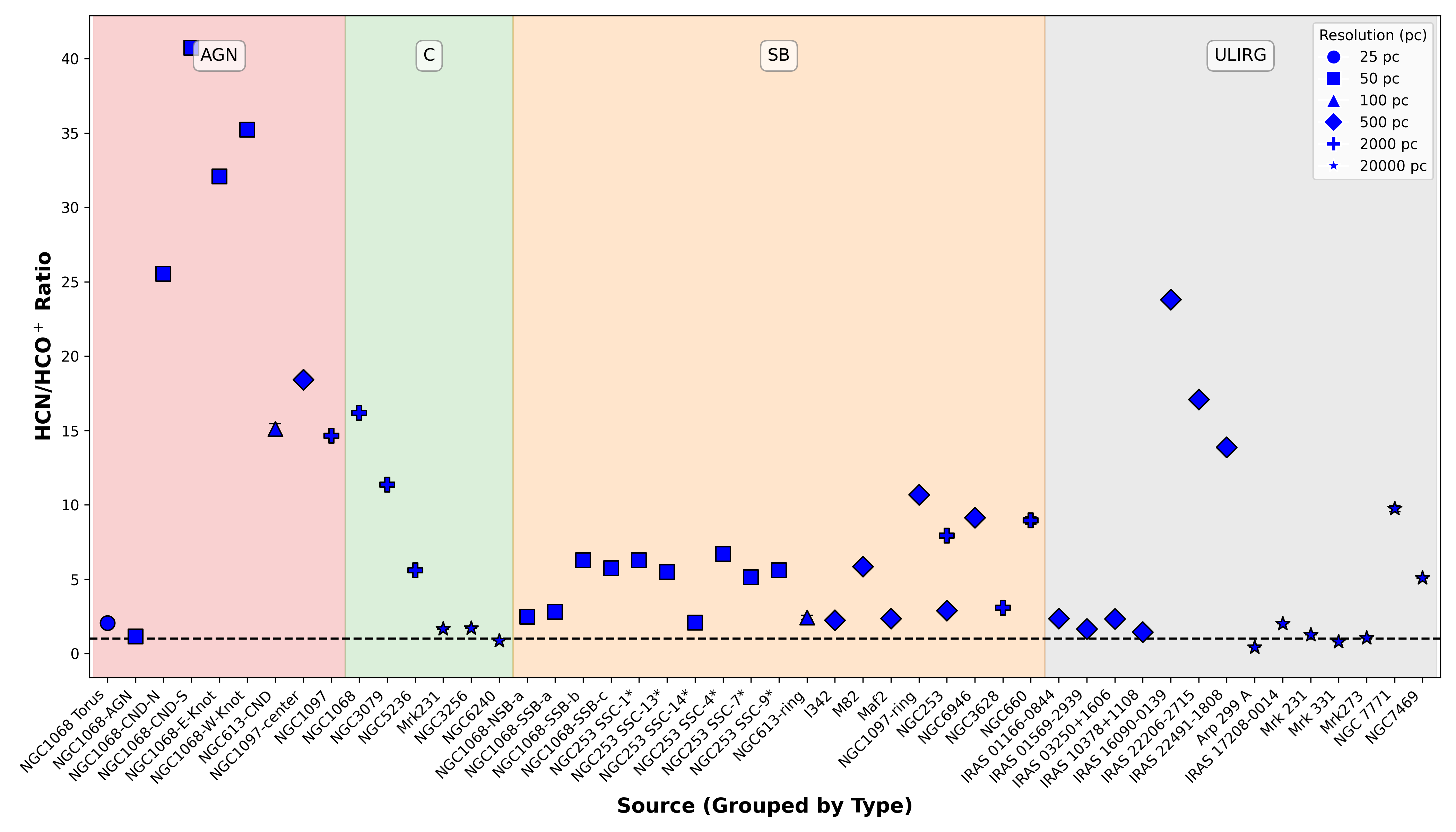}
    \caption{A swarm plot of the resulting N(HCN)/N(HCO\plus) ratios obtained by fitting the observed intensity ratios of HCN and HCO\plus to \texttt{RADEX} models. The dashed horizontal line shows the points at which N(HCN)/N(HCO\plus)=1. The shape of each point shows the resolution at which the ratios was obtained.}
    \label{fig:Model}
\end{figure*}

Figure \ref{fig:Model} presents all of the ratios produced from the resulting column densities of HCN and HCO\plus by the \texttt{RADEX} analysis. All the predicted column density results for each resolution for each source are shown together. The ($n_{\rm H2}$) and ($T_{\rm kin}$) parameters have a known degeneracy in \texttt{RADEX} models resulting from the similar excitation conditions that may occur for molecules such as HCN and HCO\plus within high-temperature, low-density and low-temperature high-density conditions \citep{Viti_2017,Butterworth2022}. As a result of this degeneracy, the focus of the discussion is upon the resulting column densities ratios. A representative posterior distribution resulting from the sampling of the CND-S region of NGC~1068 is provided in Appendix \ref{app:posterior}.

Considering each galaxy type first, we find that, generally, to explain the observed line intensities found in defined AGN-dominated systems, a high ($\geq 15$) N(HCN)/N(HCO\plus) ratio is required; this is the case for all sources, except the AGN position of NGC 1068. This source was remarked upon, particularly in relation to the NGC 1068 Torus observations that showed a higher HCN/HCO\plus intensity ratio in J(3-2) but similar in the J(4-3) ratio. In fact for HCN and HCO\plus their column density ratios within both these regions are similar, with the low column density ratio of NGC1068-AGN resulting from the low intensity ratios observed at this resolution (see Section \ref{sec:Cloud_ratios}). The low NGC 1068 Torus N(HCN)/N(HCO\plus) ratio is likely driven by the similarly relatively low J(4-3) ratio observed in this region, when compared to other regions within the CND of NGC~1068. The `composite' sources mostly comprise of low spatial resolution ($\geq2000$ pc) observations that  encompass both an AGN as well as SB regions. As such they do appear to reside in a middle ground between the AGN and SB regions. 

Our starburst group, consisting of the most discrete sources, is also the most consistent group in terms of predicted N(HCN)/N(HCO\plus) ratio, with all sources constrained to within one order of magnitude. For these sources there seems to be less impact of resolution upon predicted column density ratio. This is perhaps due to the fact that they do not experience the same `contamination' effect from AGN activity. Finally, the (U)LIRG sources appear to mostly occupy the same range of column density ratio values as the SB sources, with the exception of 3 sources, IRAS 16090-0139, IRAS 22206-2715 and IRAS 22491-1808. These sources, each observed as a part of the work done by \cite{2023_Imanishi}, are notable for their relatively high HCN/HCO\plus intensity ratios relative to the rest of the ULIRG sources. This clear enhancement of HCN relative to HCO\plus within these sources is comparable to AGN sources at the same resolution. This may imply that the enhancement of the HCN resulting from the AGN may be observed within certain AGN at $\sim 500$ pc. We note that chemical models have suggested that HCO\plus may experience an enhancement in abundance via the destruction of CO with H$_3^+$ in regions of high cosmic ray ionisation rate, such as those predicted in ULIRGS \citep{2010_HCO+_react}. The abundance of HCO\plus as influenced by varying degrees of cosmic rays has been shown by models to be highly dependent on many factors, such as density, and may even result in an anti-correlation between cosmic ray ionisation rate and HCO\plus abundance in some cases \citep{2011_Bayet,2011_Meijerink,2019_Gaches}. Nevertheless it is possible that the variations observed in the HCN/HCO\plus ratio within different ULIRGs may be contributed to by the enhancement of HCN. More observations of these sources alongside more observations of HCN and HCO\plus within ULIRGs may be necessary first before strong conclusions can be made.

\section{Summary and conclusion}
\label{sec:Conclusion}

In this paper we have used observations HCN/HCO\plus ratios across many nearby extragalactic sources that are dominated by varying active processes (AGN, star formation) at multiple spatial scales. Doing this has allowed us to more precisely defined the conditions (e.g. required spatial resolution, type of source) under which the HCN/HCO\plus ratio may be used as a tracer of AGN or SB activity. Furthermore, we have also investigated whether radiative transfer modelling may be used in order to further distinguish between types of sources where intensity ratios cannot. Our main conclusions are as follows:

\begin{itemize}
    \item The intensity ratios of HCN/HCO\plus appears to be able to trace AGN activity vs starburst activity in approximately cloud scale resolution (25-50pc). At lower resolutions,($>100$pc),  however, contamination from multiple emission sources leads to an evident stochastic nature of this  ratio. This is particularly prominent at extremely low resolution observations that are approximately galactic-scale.
    \item Of the HCN/HCO\plus ratios higher J-transitions, such as J=3-2 and J=4-3, are favoured. Ratios of these transitions show the greatest trends away from unity, above in the case of AGN-dominated regions, and below in the case of SB-dominated. This is consistent with the conclusions of \cite{Butterworth2022} that found HCN(4-3)/HCO\plus(4-3) to be a significantly better tracer than HCN(1-0)/HCO\plus(1-0) in NGC~1068.
    \item Radiative transfer models were used to interpret the likely causes for the observed increase of HCN/HCO\plus within AGN-dominated regions. As has been proposed and supported by previous studies the likely cause is an enhancement in HCN abundance relative to HCO\plus \citep{Izumi2016,Imanishi_2018}. This theory is supported by our models. The AGN-dominated sources (with the exceptions located at the center of NGC1068) all appear to show notable enhancements of HCN when compared to the SB sources.
    \item ULIRG sources with multiple transitions of HCN and HCO\plus show a similar trend of column densities as to that of SB sources. Given the fact that there are some exceptions to this, this means that tracing the AGN activity of ULIRGs may be possible as long as one uses modelling. Consideration however, must be given to other external properties likely to affect the HCN/HCO\plus ratio within ULIRGs.
\end{itemize}

A further expansion of the number of same-J observations of HCN and HCO\plus across nearby galaxies at the high resolutions available thanks to modern interferometers can only further inform us of the processes that produce the observed enhancement of HCN in AGN-dominated regions. Furthermore, our understanding of why this relation may further confirm whether it may, or may not, be used for galaxies like ULIRGs and/or whether should be restricted to cloud-scale active regions.

\section*{Data Availability}
 A table of the data used throughout this paper is only available in electronic form at the CDS via anonymous ftp to \href{http://cdsarc.u-strasbg.fr/}{cdsarc.u-strasbg.fr} (130.79.128.5) or via \href{http://cdsweb.u-strasbg.fr/cgi-bin/qcat?J/A+A/}{http://cdsweb.u-strasbg.fr/cgi-bin/qcat?J/A+A/}.

\begin{acknowledgements}
This work is part of a project that has received funding from the European Research Council (ERC) under the European Union’s Horizon 2020 research and innovation programme MOPPEX 833460. This research has made use of the NASA/IPAC Extragalactic Database (NED), which is funded by the National Aeronautics and Space Administration and operated by the California Institute of Technology. The published data displayed in the following papers were used in this study: 1 - \cite{2017_Sliwa}; 2 - \cite{2008_gracia-Carpio}; 3 - \cite{2023_Israel}; 4 - \cite{2019_Kawamuro}; 5 - \cite{2022_Sato}; 6 - \cite{2022_Tristram}; 7 - \cite{2024_Nishimura}; 8 - \cite{Tan2018}; 9 - \cite{2023_Imanishi}; 10 - \cite{Aladro2015}; 11 - \cite{Krips_2008}; 12 - \cite{Butterworth2022}; 13 - \cite{GarciaBurillo_S_2008}; 14 - \cite{2019_Garcia_Burillo}; 15 - \cite{Imanishi_2020}; 16 - \cite{2012_Hsieh}; 17 - \cite{Izumi2013}; 18 - \cite{2013_Imanishi}; 19 - \cite{2021_Ueda}; 20 - \cite{2021_Audibert}; 21 - \cite{2024Butterworth}; 22 - \cite{Meier_2015}; 23 - \cite{2017_Jimenex-donaire}; 24 - \cite{2014_Zhang}; 25 - \cite{Audibert_2019}; 26 - \cite{Miyamoto_2017}; 27 - \cite{2015_Izumi}. 
\end{acknowledgements}

%
%
\bibliographystyle{aa}
\bibliography{refs} 


\clearpage
\appendix

\onecolumn

\FloatBarrier
\section{The scale of Observations of NGC~1068}
\label{app:1068}
\begin{figure*}[h!]
    
    \includegraphics[width=\linewidth]{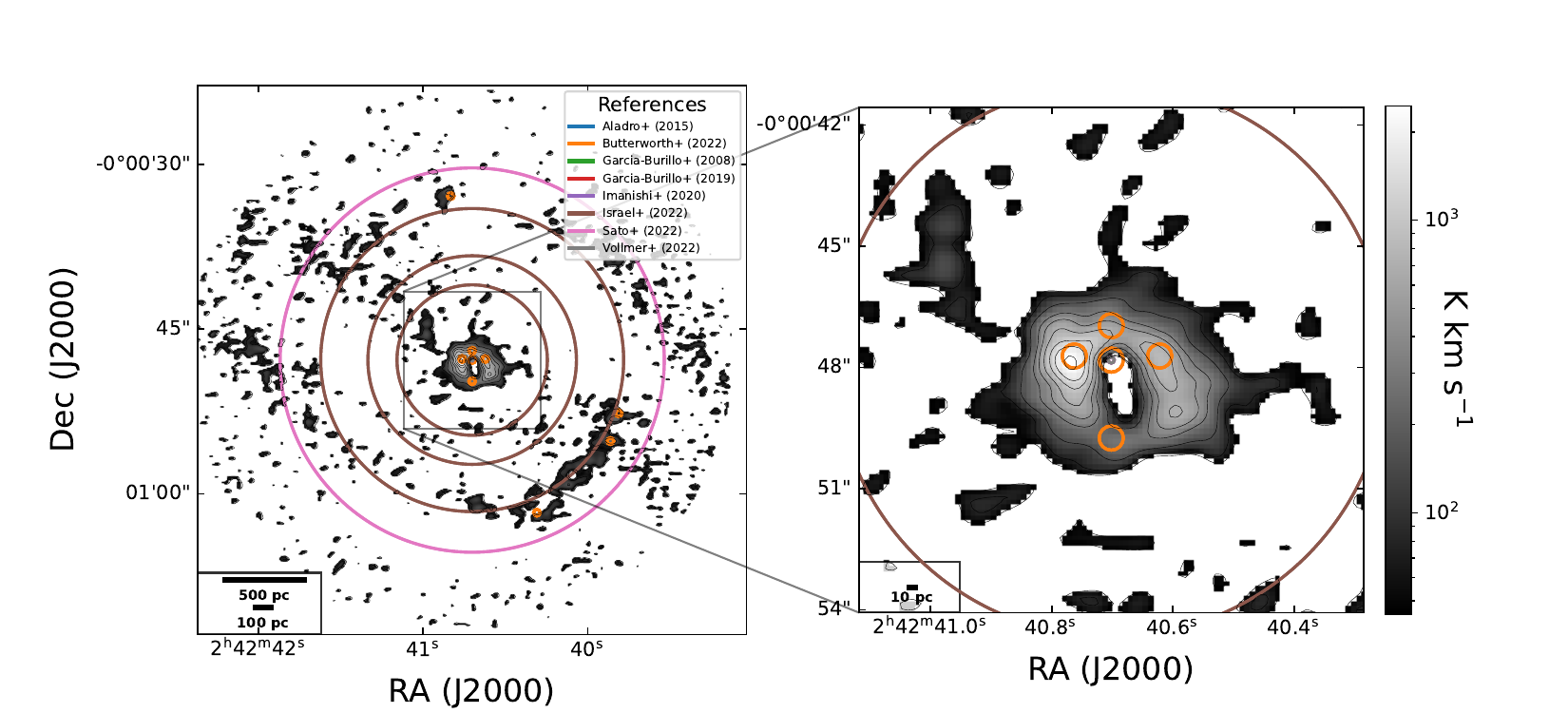}
    \caption{A visual representation of the approximate scales and regions covered by observations of the composite galaxy NGC~1068 as used within this study plotted over the velocity integrated moment 0 map of HCN~(1-0) with a 3$\sigma$ cut off.}
    \label{fig:posterior}

\end{figure*}

\newpage
\FloatBarrier
\section{The Posterior Distribution of the CND-S of NGC~1068}
\label{app:posterior}
\begin{figure*}[h!]
    
    \includegraphics[width=\linewidth]{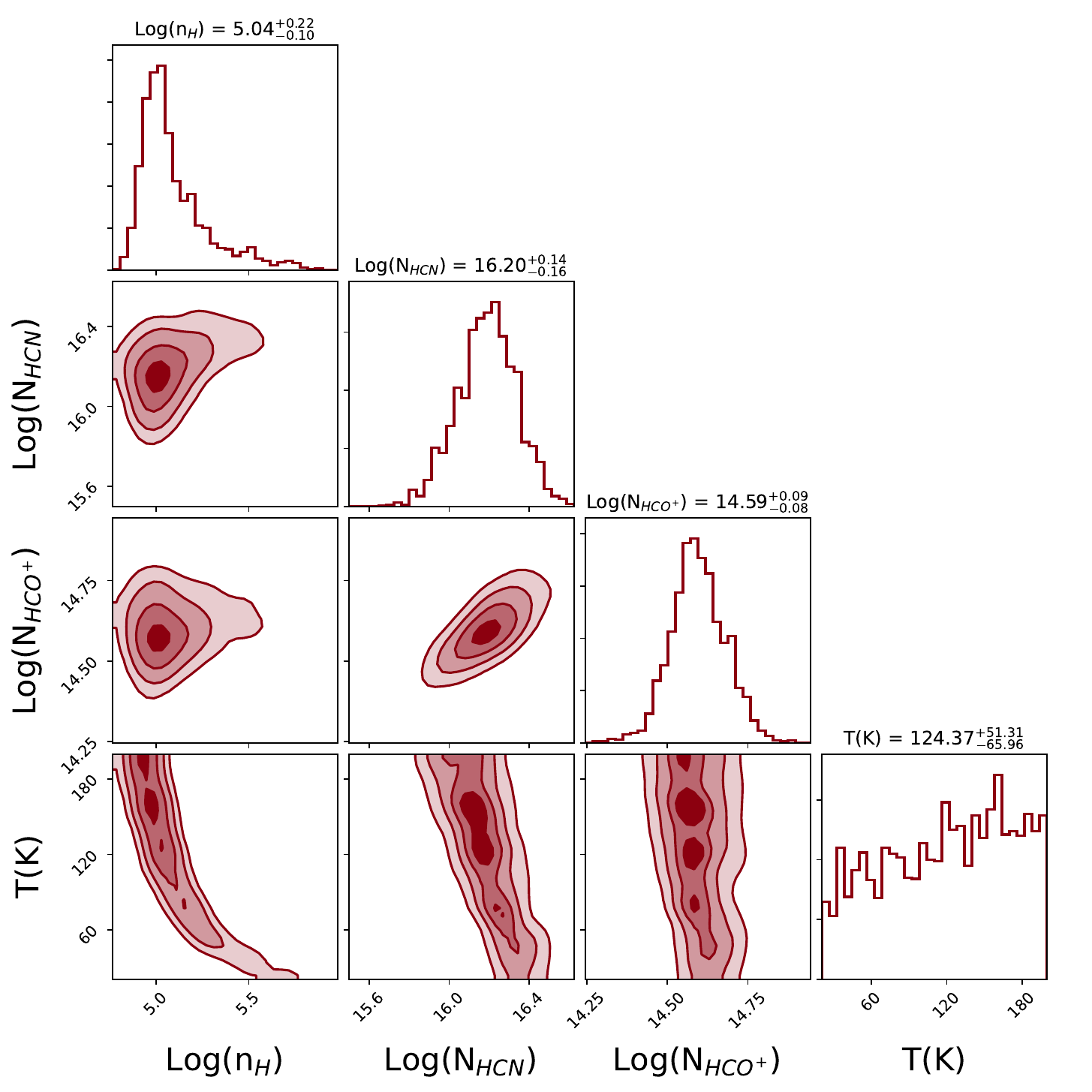}
    \caption{The resulting posterior distribution of the CND-S region of NGC~1068 as defined in \cite{Butterworth2022} across the 4 parameters of Kinetic temperature, neutral gas density and column densities of HCN and HCO\plus.}
    \label{fig:posterior}

\end{figure*}

\end{document}